\documentclass[sigconf]{cidr-2025}
\usepackage{graphicx}
\usepackage{subcaption}
\usepackage{multirow}

\usepackage{listings}
\usepackage{xcolor}
\usepackage{verbatim}

\definecolor{codegreen}{rgb}{0,0.6,0}
\definecolor{codegray}{rgb}{0.5,0.5,0.5}
\definecolor{codepurple}{rgb}{0.58,0,0.82}
\definecolor{backcolour}{rgb}{0.95,0.95,0.92}

\lstdefinestyle{mystyle}{
    backgroundcolor=\color{backcolour},   
    commentstyle=\color{codegreen},
    keywordstyle=\color{magenta},
    numberstyle=\tiny\color{codegray},
    stringstyle=\color{codepurple},
    basicstyle=\ttfamily\footnotesize,
    breakatwhitespace=false,         
    breaklines=true,                 
    captionpos=b,                    
    keepspaces=true,                 
    numbers=left,                    
    numbersep=5pt,                  
    showspaces=false,                
    showstringspaces=false,
    showtabs=false,                  
    tabsize=2
}

\usepackage{soul}
\usepackage{todonotes}
\presetkeys{todonotes}{inline}{}

\AtBeginDocument{%
  \providecommand\BibTeX{{%
    \normalfont B\kern-0.5em{\scshape i\kern-0.25em b}\kern-0.8em\TeX}}}

\begin{document}

\title{Beyond Relations: A Case for Elevating to the Entity-Relationship Abstraction}

\author{Amol Deshpande}
\email{amol@umd.edu}
\affiliation{%
\institution{University of Maryland}
\city{College Park}
\country{USA}
}

\renewcommand{\shortauthors}{Deshpande}

\begin{abstract}
Spurred by a number of recent trends, we make the case that the relational database systems should urgently move beyond supporting the basic object-relational model and instead embrace a more abstract data model, specifically, the entity-relationship model. We argue that the current RDBMSs don't inherently support sufficient {\em logical} data independence, and that is relegating the database systems to the role of
a backend storage system, away from where significant innovation is both happening and is still needed. We present the design of a prototype system ({\em ErbiumDB}) that we are building to explore these issues, and discuss some of the key research challenges.
\end{abstract}

\maketitle

\section{Introduction}

The relational model, proposed over 50 years ago, has been the foundation of most high-performance database systems to date. The object-relational model, i.e., the relational model with abstract data types and related functionality, has generally satisfied the needs of most enterprise applications without being overly complicated. 
A number of higher-level {\em semantic data models} have been proposed over the
years~\cite{peckham1988semantic}, including by Ted Codd himself~\cite{codd1979extending}, to better support
complex data types, inheritance hierarchies, richer integrity constraints, etc. However, none have gained significant traction although several have 
significantly influenced the refinements to the relational
model~\cite{zaniolo1983database,hammer1978semantic} (we cover some of this work in Section~\ref{sec:history}). As articulated by Stonebraker et al.~\cite{stonebraker2005goes,stonebraker2024goes}, this was primarily due to the newer data models not offering significant improvements or benefits to users, and also because the most important features were easy to incorporate into the relational model.

\subsection{Why Now?}
However, we argue that database systems are at a pivotal juncture and they need to decide whether to raise the
abstraction level offered by them and come closer to the users and the application developers. Failing that, we believe
that database systems will be increasingly relegated to serve as backend storage systems, with most of the application
logic and user interactions served by layers on top. 
We won't rehash all the reasons that have been laid forth in the past for
richer data models, but make a few observations based on recent developments:

\vspace{6pt}
\noindent\textbf{(1) \underline{Unwieldy and hard-to-understand schemas:}} First, relational schemas in practice are often large, unintuitive, and hard to understand, making it difficult for new or even experienced users to construct and verify SQL queries for a given task. The RDBMSs often lack sufficient context around the attributes, and documentations explaining the schema (usually maintained externally) are rarely kept up-to-date; foreign keys and other constraints, intended to help incorporate structure into the schema, are often not used properly and systematically. 
A large part of the reason for the above is \textbf{schema ``decay''}~\cite{stonebraker2016database} arising out of small
incremental changes made to the original normalized schema (often constructed from an E/R diagram). The basic relational
model is not sufficiently ``opinionated'' and easily permits schema changes that, over time, allow the schema to stray
far from the principled normal forms. As a separate challenge, database systems today do not support {\bf schema evolution}
natively~\cite{curino2008graceful,caruccio2016synchronization,bhattacherjee2021bullfrog}, requiring users to build significant scaffolding on top to handle schema changes and resulting {\em data migrations}.

Increased use of large language models for coding and data analysis will likely make this problem more challenging as
the focus increasingly shifts toward understanding SQL queries rather than crafting them. It is difficult to understand
and verify multi-way join queries over a large number of tables typically found in real-world databases. Higher-level abstractions 
can help mitigate this problem, both for the LLMs and the users~\cite{zheng2023take}.

\vspace{4pt}
\noindent\textbf{(2) \underline{Data governance and compliance:}} Second, data governance and compliance issues are an increasing concern for many enterprises.
Although it took a back seat over the last few years, compliance with privacy or AI regulations like GDPR, CCPA, etc.,
increasingly requires a more careful accounting and handling of personal data~\cite{deshpande2021sypse,angin2010entity,makhdoom2024securing}. In addition to better understanding 
and tagging the data being collected, compliance 
often also requires fine-grained access control and ability to delete data of specific individuals, both of which are
fundamentally {\bf entity-centric operations}, i.e., operations that require reasoning about all the data related to an entity (a person or an organization) as a whole. These tasks 
are challenging to do in a verifiable manner for normalized relational schemas where personal data may be spread across many tables, often without the foreign keys to help link the data. 
The common workarounds today include external metadata managers that typically support more user-friendly abstractions, but are difficult to set up and maintain, and are not sufficiently 
integrated with the database or the application code.

\vspace{4pt}
\noindent\textbf{(3) \underline{Impedance mismatch:}} Third, due to the impedance mismatch between the relational model and the entity-centric nature of
application code, most applications use some sort of a data layer on top of the database, typically an object-relational mapper (ORM) or an API service. 
The ORMs typically expose an abstraction that is very similar to the entity-relationship model, often borrowing the same terminology~\cite{o2008object}; whereas the API services typically 
expose hierarchical views on top of the database. In either case, the intermediate layers are responsible for translating the higher-level abstractions (e.g., GraphQL) 
into SQL. This not only leads to significant duplication of effort, but the mappings across the layers can be difficult to maintain and can result in data inconsistencies~\cite{bailis2015feral}. 
Furthermore, most RDBMSs today support much of the required functionality, including the ability to generate nested outputs in the SELECT clause (e.g., PostGraphile\footnote{\url{https://www.graphile.org/postgraphile/performance/}} can compile a highly nested GraphQL query into a single PostgreSQL query).

\vspace{4pt}\noindent\textbf{(4) \underline{Adoption of JSON and Property Graph Models:}} For similar reasons, other data models like documents (XML, JSON), property graphs, RDF, have seen much work and some commercial success over the last decade. Although there exists workloads that are specifically suited for those models (e.g., graph analytics tasks such as community detection), in most cases, the data and query workloads for those databases are very similar to relational database
workloads; in several cases, these databases are often built as layers on top of a relational database. In our opinion,
unifying these models is a pressing challenge for our community, as it has led to significant and unnecessary fragmentation and duplication of effort, both on the user-facing end as well as the backend storage and query execution layers. 

\vspace{4pt}\noindent\textbf{(5) \underline{Lack of logical independence:}} Finally, we are leaving a lot of performance on the table by having the user-facing abstraction so closely tied to the physical layout of the data. Relational databases have exploited physical data independence to dramatically improve performance and adapt to evolving hardware patterns as well as parallel and distributed environments. However, the low-level nature of the relational abstraction 
leads to much lower logical independence. Decisions about how to represent relationships, what to do with multi-valued attributes, how to handle inheritance, etc., need to made when designing the schema, and can have a significant impact on the performance. These decisions can and should be {\em auto-tuned}, i.e., made based on the workload patterns and requirements. In theory, {\em views} can be used to 
achieve logical data independence, but require significant expertise to use correctly and it can be difficult to optimize queries in their presence.
We present several illustrative experiments in Section~\ref{sec:experiments} to highlight the potential performance benefits of the increased logical data independence.

\subsection{Why E/R?}
Given these reasons, we argue that more effort should be spent in desiging and supporting a higher-level abstraction
natively in the database systems. 
We specifically advocate for
the familiar (extended) {\em entity-relationship} abstraction~\cite{chen1976entity}, which 
is already widely used for inital schema design and may be considered richer than document models or the property graph model for 
most use cases (assuming schemas are enforced). Similar to the document models, the E/R model supports fixed-depth hierarchical/nested data (through the use of composite attributes) as well as arrays (through the use of multi-valued attributes); it also inherently supports {\em relationships} which is a key missing element in the document data model. E/R model does require a more rigid schema design and doesn't support arbitrary nesting. However, schema-less or schema-flexible designs shift the burden of managing the schema to the developer, moving significant complexity and business logic (e.g., constraints) to the application code. To the extent that those features are actually used in practice, they can be supported through the use of a JSON type. At the same time, we believe that the E/R model lends itself to easier {\em schema evolution and management}, mitigating the need for schema flexibility that is often cited as a reason for using those systems.

Property graph databases with schemas also naturally map to the E/R model; in fact, the recent proposal for property graph schemas~\cite{angles2023pg} uses the E/R abstraction as a starting point and we don't see strong distinctions between the two (the distinctions mentioned in related work often trace back to incomplete development of the E/R model as a practical data model). Similarly, although we use a relational backend (PostgreSQL) in our prototype, a property graph database system is perhaps a more natural backend alternative for the E/R model. 

We are cognizant of many failed {\em now-is-the-time} attempts to move away from the relational model in the past.
However, we believe a large part of that may be a ``self-fulfiling prophecy''. There hasn't been a concerted effort to provide a combination of a more abstract
data model and high performance, so the users are left with choosing between the two (and they often choose the more
user-friendly model first, and reluctantly switch to the latter when scale becomes an issue). We point to the prevalent
and increasing use, over the last decade or two, of:\\[2pt]
(a) ORMs for web development, \\[2pt]
(b) API services (e.g.,
PostgREST\footnote{https://docs.postgrest.org/en/v12/}), \\[2pt] (c) document, property graph, or
graph-relational\footnote{\url{https://www.edgedb.com/blog/the-graph-relational-database-defined}} models, \\[2pt]
(d) hierarchical storage formats like Parquet and Avro that have become the dominant
storage formats in data lakes, and  \\[2pt]
(e) query and analysis engines like Apache Datafusion or DuckDB that provide SQL-like interfaces on top of those formats. 

\begin{figure*}[h]
  \centering
  \includegraphics[width=\linewidth]{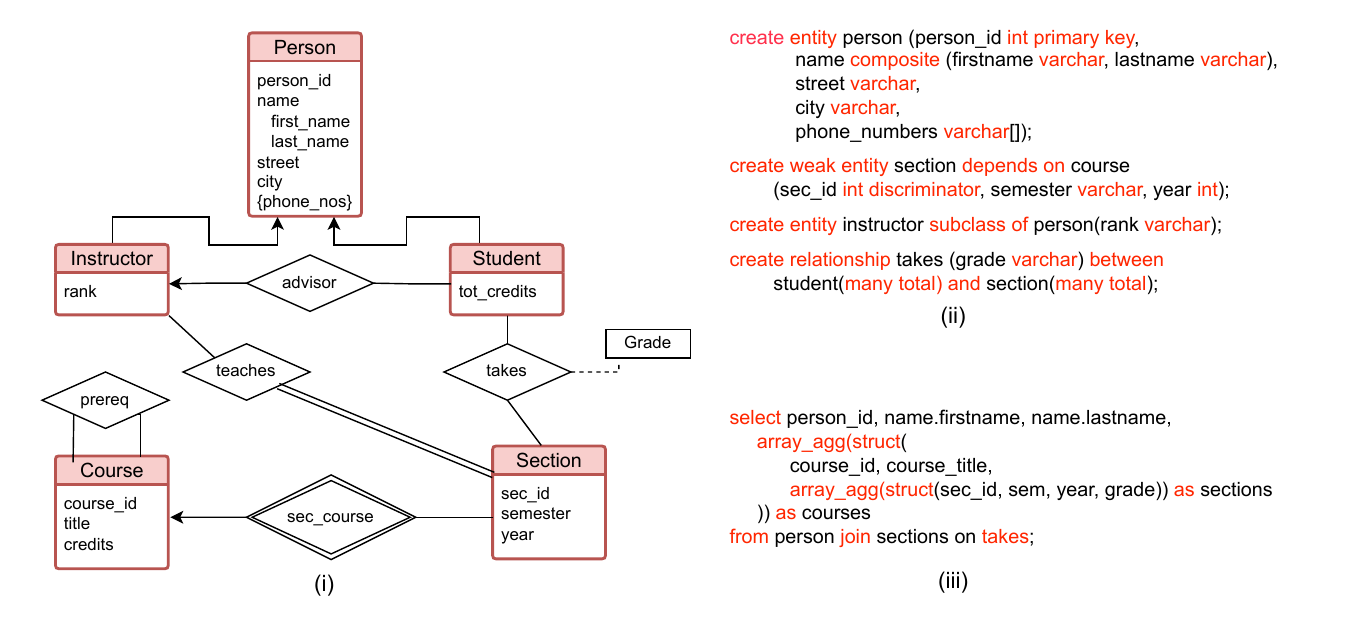}
  \vspace{-12pt}
  \caption{(i) Example of an E/R model with one weak entity set and two subclasses (adapted from~\cite{silberschatz2019database}); (ii) DDL to create entities and relationships;
(iii) An example query for illustration purposes}
  \vspace{-5pt}
 \label{fig:er_example}
\end{figure*}

Performance is often cited as a reason to avoid higher-level abstractions. Drawing an analogy to the 70s when similar criticisms were made of the relational model, we believe that the increased optimization opportunities due to the higher-level abstraction will ultimately lead to better performance. There are a number of engineering challenges that we would need to address, but we don't see any fundamental reasons why those would be insurmountable. 

On the OLTP side, two of the key challenges are: (a) handling complex objects with nesting and arrays, and (b) possibility that a single update may require updating multiple tables (depending on the mapping of the E/R model to the physical storage). For the former, serialization formats like protocol buffers support nesting and arrays, and can be used for client-server communication without additional overhead, whereas binary formats like BSON are already widely used for storage. The latter challenge may require development of new transaction processing techniques, and may not be achievable by building on top of an existing system (as we do in our prototype at this time). For OLAP workloads, data warehouses or lakehouses already support hierarchical storage formats, and the success of Parquet, DataFusion, etc., suggest that the performance issues can be easily ameliorated. However, inheritance hierarchies pose a major challenge as they may result in a large number of left outer joins~\cite{melnik2008compiling} if the E/R model is implemented on top of a relational database (or we suspect property graph databases). We believe this necessitates development of new storage layouts and query processing techniques, and forms one of the more interesting research questions in this area.

To explore these challenges in more depth, we are building a prototype that supports the entity-relationship model as
the primary data model, and an SQL-like query language against that model; we
share some of the reservations about SQL from recent work~\cite{carey2024sql++,neumann2024critique,shute2024sql}, and believe that a functional transformation-based query language 
is easier to 
understand and offers more flexibility and extensibility, especially when working with richer data models, and that queries written in those likely require fewer changes when 
schemas are modified. However, we leave a deeper investigation of that to future work.

In the rest of the paper, we elaborate on some of the design aspects of the system, and discuss the spectrum of storage formats that we believe should be supported in the backend
(including a compressed multi-relation representation). We then present
a systematic way to explore the optimization opportunities enabled by the increased logical data independence. We also plan to support schema evolution and versioning natively in
our system, but we omit a detailed discussion since that's not the focus of this paper.

\section{Abstractions}
We propose using the standard (extended) E/R model as the starting point due to its familiarity as well as widespread use as the conceptual model for schema design. 
It is also aligned with the abstraction supported by many ORMs (e.g., Django) and similar systems like EdgeDB. 
Since the E/R model supports modeling of hierarchical information (through composite and multi-valued attributes), it also naturally encompasses the key elements
of the more flexible models like property graphs and hierarchical documents, as discussed earlier. 
E/R model doesn't inherently support the notion of an ordered list, but that can either
be added explicitly or handled through the use of a positional attribute as needed.

Figure \ref{fig:er_example} shows an example of an E/R model (adapted from the running example from~\cite{silberschatz2019database}), that contains one {\em weak entity set}, two 
subclasses ({\em specialization}), and various types of relationships (with annotations to indicate the cardinality and participation constraints). The figure also shows the syntax for defining the entities and relationships in our prototype, including the 
ability to directly define composite attributes (which would require the use of a separate type in a typical RDBMS) as well as multi-valued attributes (technically not allowed in the
relational model, but supported by most RDBMSs today). In addition, the DDL should support: (a) defining
{\em constraints}, (b) specifying the inheritance properties (total vs partial, disjoint vs overlapping, etc.), and (c) adding descriptive text (that can be automatically used, e.g., for creating API documentations).  There is much work on (a) and (b) that we plan to build upon.

For querying, we use a variant of SQL with an example shown in Figure \ref{fig:er_example}. The two main additions to normal SQL that we support:
\begin{itemize}
\item Ability to specify a relationship when joining two relations, in addition to standard WHERE and ON clauses. 
\item Ability to construct hierarchical outputs in the SELECT clause. We borrow Apache DataFusion's syntax for this purpose, however other constructs (e.g., SQL++~\cite{carey2024sql++}) can also be used instead. This is a very common use case in practice, and although
it can be supported through a chain of \verb|array_agg| and \verb|group by|'s, we believe this functionality should be supported natively so that the queries can be optimized 
properly. We also omit explicit \verb|group by| clauses as shown in the example, since those can be inferred from the
\verb|select| clause (otherwise CTEs need to be used to construct the output shown here).
\end{itemize}

\section{Schema Evolution}
In addition to making data governance tasks easier, a key motivation for us is better support for {\em schema evolution}.
Stonebraker et al.~\cite{stonebraker2016database} lay out many reasons why schemas in the wild become unmanageable, unnormalized, and stray from the 
original E/R diagram. They also explore, but ultimately dismiss, the idea of raising the abstraction level to something like an E/R model, claiming that 
it would not solve the problems. We disagree with that conclusion, and believe that database decay will be at least delayed if a higher-level data model is used, 
especially if shortcuts are not allowed (i.e., the database is more ``opinionated'').

For instance, consider a schema change where a single-valued attribute is made multi-valued (e.g., moving from a single city to multiple cities).
In a normalized relational schema, this requires moving the attribute to a separate table, and any queries that access that attribute would need to be modified
to do an additional join. However, this is a minor change for the E/R diagram, and results in relatively localized changes to queries involving that attribute 
(e.g., \verb|select person_id, city| $\rightarrow$ \verb|select person_id, unnest(city)|). Note that, internally we may wish to store the attribute separately, 
but that decision can be made transparently to the user.

Similarly, converting a many-to-one relationship to a many-to-many relationship requires creation of a new table (typically) in the relational schema and queries
need to be modified appropriately. The change to E/R diagram is again relatively minor, and queries involving the relationship may not need any modifications 
depending on what it was trying to achieve. For example, a query to find average credits per advisee for each
instructor: \\[3pt]
\verb|  select instructor.ID, avg(tot_credits)|\\
\verb|    from instructor join student on advisor|\\
\verb|    group by instructor.ID| \\[2pt]
does not require any modifications if the relationship cardinalities were to be modified. There are of course other queries where changes may be needed.

\begin{figure*}[h]
  \centering
  \includegraphics[width=\linewidth]{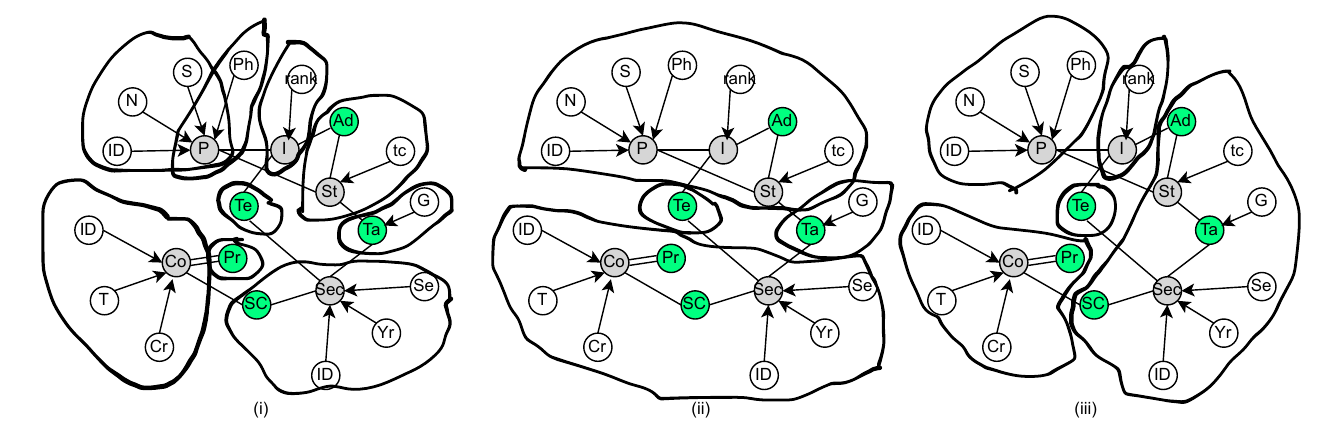}
\vspace{-5pt}
  \caption{Mappings to physical representation as covers of the E/R graph}
\vspace{-5pt}
 \label{fig:mappings}
\end{figure*}

A major advantage of the E/R model is the ability to define class hierarchies like the one shown in Figure \ref{fig:er_example}. When constructing
a relational schema from such an E/R diagram, we need to choose between multiple possibilities depending on whether the specialization is total vs partial, and 
whether it is overlapping or disjoint, etc. There are at least three possibilities here: 
\begin{itemize}
\item Three relations, \verb|Person|, \verb|Instructor|, and \verb|Student|, each storing a disjoint set of individuals, with the latter two featuring an extra attribute each.
\item Three relations, \verb|Person|, \verb|Instructor|, and \verb|Student|, but \verb|Person| stores all the common attributes for all individuals, and the other two
relations only store the additional attributes (along with the key).
\item A single relation \verb|Person|, with \verb|rank| and \verb|tot_credits| appropriately set to null (or with
explicit attributes to keep track of who belongs to which subclass).
\end{itemize}
The choice among these options needs to be made early on, and gets baked into any queries that are written against this schema. Switching between options would likely 
require significant modifications to most queries. 
This is a less drastic change for the E/R model, although semantic correctness can still be an issue for some queries.
This choice can also have a significant impact on performance depending on the workload, something that is not easy to anticipate a priori. 

Finally, we note that schema changes typically also require a complex {\bf data migration} process, which today is often handled by the application layers on top since
databases do not support such functionality natively. We plan to explore how schema evolution and data migration can be supported natively within the database system, 
along with {\em versioning}~\cite{bhattacherjee2015principles}, so that users can more easily experiment with schema changes and roll them back as needed. Although these are orthogonal
issues, we believe the use of a higher-level model simplifies some of the challenges considerably.

\section{Mapping to Physical Representation}
The logical independence afforded by the higher abstraction layer enables a vastly broader space of physical
representations that can be used to store the data on the underlying storage system. Before describing our approach, we briefly discuss some of the prior work
along these lines. In the initial C-store prototype~\cite{cstore}, a
larger space of physical representations was considered in the form of {\em projections}. Specifically, a given
relational schema could be mapped to a set of projections that do not necessarily correspond one-to-one or even
many-to-one 
with the tables in the schema. A projection could contain attributes from multiple tables connected through
appropriate join keys and could be sorted and stored differently. However, to our knowledge, later work on columnar 
databases considers a much narrower set of projections. Join Indexes~\cite{valduriez1987join} similarly expand
the space of physical representations considered, however, those are typically restricted to join keys. 
The work on XML, RDF, or property graph ``shredding''~\cite{tatarinov2002storing,neumann2010rdf,sun2015sqlgraph} has
explored different ways to map from those models to relatioal. The target of most of that work is, however, tabular
representation, whereas our approach explores a larger space of physical representations including hierarchical
representations. Another closely related work is the line of work on converting from E/R model to
XML~\cite{franceschet2013graph,elmasri2005conceptual}, where the target is (effectively) a set of hierarchical representations. That work however, hasn't
looked at systematic exploration of the space of possible mappings in a workload-aware fashion (instead focusing on finding the best representation 
given an optimization metric), and also doesn't consider multi-relation representations.
Finally, the ADO.NET Entity SQL framework~\cite{castro2007ado,melnik2008compiling,rull2013query} allows users to write queries against an E/R-like conceptual
model; the relationship between that and the storage backend is specified using a declarative mapping, that is compiled into {\em bidirectional views} that are
used to transform data back and forth. We plan to build upon that considerable line of research in our future work.

At a high level, the goal of the mapping optimization process is to create a collection of physical representations that can be used to store the 
data that conforms to the given E/R schema. There are two key requirements: \\[2pt](1) The mapping must be uniquely reversible (i.e., bidirectional)
in that, the entities and relationships stored in the database must be recoverable, and\\[2pt] (2) We must be able to map any inserts/updates/deletes 
to the entities and relationships to the database. 

We currently consider three possible physical representations that can be used as targets.
\begin{itemize}
\item {\bf Tables in the first normal form,} where composite data types are permitted, but domains must be atomic (i.e., no
arrays). This was typically the only representation considered by the prior work on shredding.
\item {\bf Hierarchical structures with a pre-defined schema:} Here we allow for arrays, including arrays of composite types
which themselves might contain arrays. Although relational databases are typically not optimized for this scenario, the
columnar storage formats like Parquet and Avro have shown that read-only workloads can be supported efficiently for such
data. However, updates are typically harder to do for such storage structures. We note that some of these issues have
been investigated in the work on nested relational databases as well~\cite{roth1988extended}.
\item {\bf Multi-relational compressed ({\em factorized}) representations:} This representation, in theory, can be used to store 
the join of multiple relations together in a compact fashion~\cite{olteanu2016factorized} (a materialized view stored as
a table, on the other hand, may have significant duplication). The key benefit here is the ability to use physical
pointers to avoid joins, and to execute some types of aggregate queries more efficiently (by, in effect, pushing
down aggregations through the joins). We expect that the benefits of this representation will likely show up if the 
joins are almost one-to-one (i.e., the join is not a key-foreign key join at the schema level, but the data does not
exhibit high connectivity). Another key reason for us to consider this is that, it brings us closer to the storage
formats used in graph databases, and thus helps unify the representations.
\end{itemize}

In order to explore the space of possible mappings, we first view the E/R diagram as a graph where each entity,
relationship, and attribute is a separate node (Figure \ref{fig:mappings}). Entity nodes are connected to the
relationships in which they participate, to subclasses or superclasses, and to their attributes. 
A mapping to physical storage representation can be seen as a {\bf cover} of this graph using {\bf connected subgraphs}. 
Each connected subgraph corresponds to a physical table or data structure, and together all of these constitute the full
physical representation.

We show three examples in Figure \ref{fig:mappings}. The first mapping depicts a fully normalized mapping, where each
entity gets its own table, many-to-one relationships are folded into the many side, and many-to-many relationships have
their own tables. For instance, the \verb|advisor| relationship is folded into the \verb|student| table, whereas
\verb|takes| and \verb|teaches| are in separate tables. The two subclasses of \verb|Person| are also in separate tables,
but only with the attributes that are unique to them. Finally the multi-valued attribute (\verb|Ph|) is stored in a
separate table (along with the key). 

The second mapping, on the other hand, reduces the number of data structures needed by combining \verb|Person| and its
subclasses into the same table, as well as using an {\em array} to store the multi-valued attribute. It also moves 
\verb|Section|s into the \verb|Course| table as an array of a composite type. Although it reduces the number of joins
required, most queries involving \verb|section| entities will require use of an {\bf unnest} operation, which is often
not optimized in modern RDBMSs (although platforms like Apache DataFusion do a better job at it).

Finally, the third mapping illustrates a scenario where two entities with a many-to-many relationship (\verb|section|
and \verb|student|) are stored in a single physical data structure. In our current implementation that is based on
PostgreSQL, this would result in significant duplication of data and also increase the cost of inserts/updates/deletes.
One of our key research goals is to explore alternative representations and their benefits.

We note that our approach is flexible enough to cover column decomposition representations, and it also allows for the
same attribute to be present in multiple data structures. Also, although it allows for attributes from multiple
relations to be in the same data structure, it is an open question as to whether it can cover the entire scope of
projections proposed for C-Store~\cite{cstore}.

The natural optimization problem here, that forms one of the key research challenges, is to automatically identify the best 
mapping for a given schema and data and query workload. A sub-question there is, how to generate such mappings in an
automated fashion so that one can search through them to make the optimization decisions. As noted above, any mapping
must satisfy the requirement that the mapping is reversible and CRUD operations are well-defined. 
Coming up with a complete and sound list of constraints on the graph cover is an interesting direction for future work,
which has overlaps with the work on updatability of views and answering queries over them~\cite{melnik2008compiling}.

\begin{figure}
  \centering
  \includegraphics[width=.9\linewidth]{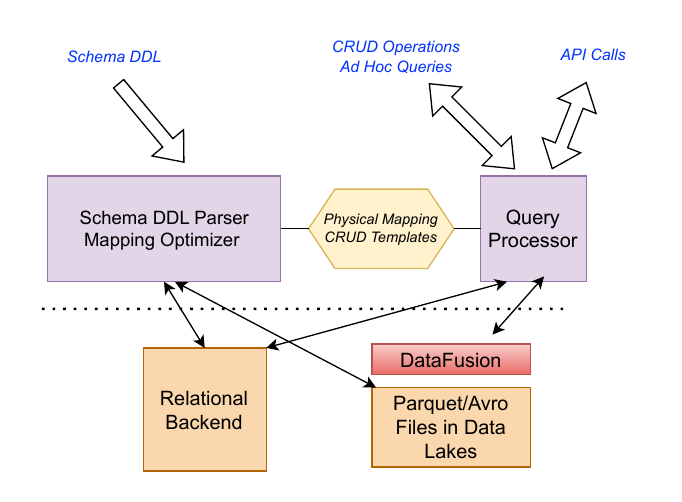}
\vspace{-5pt}
  \caption{High-level ErbiumDB Architecture}
\vspace{-5pt}
 \label{fig:arch}
\end{figure}

\section{Prototype}
We are building a proof-of-concept system, called ErbiumDB\footnote{\url{https://github.com/umddb/ErbiumDB}}, to explore the research questions raised in this paper. Our
prototype is written in Python, and built as a layer on top of PostgreSQL (with the goal to support other backends like
Apache DataFusion). Figure \ref{fig:arch} shows a high-level architecture (not all pieces are built as yet). The DDL
layer does the heavy lifting here. It creates the E/R graph from the entity/relationship \verb|create| statements, and
keeps it up to date as the schema is modified. The mapping of the E/R graph to physical tables (specified manually today) is maintained in a table in
the database as a JSON object, and is read into memory at initialization time. As noted earlier, we construct a table
for each connected subgraph in the mapping that is chosen. The DDL layer also constructs mappings between CRUD
statements on the entities/relationships to updates on the physical tables in the database. The prototype supports a
limited form of SQL, and the queries against the logical schema are translated to queries against the physical tables.
Finally, we are also planning to support a RESTful API by default (and possibly gRPC), to ensure compatibility with standard application 
development practices. While CRUD operations would be supported by default, additional API calls can be added as needed.

\begin{figure}[t]
  \centering
  \includegraphics[width=\linewidth]{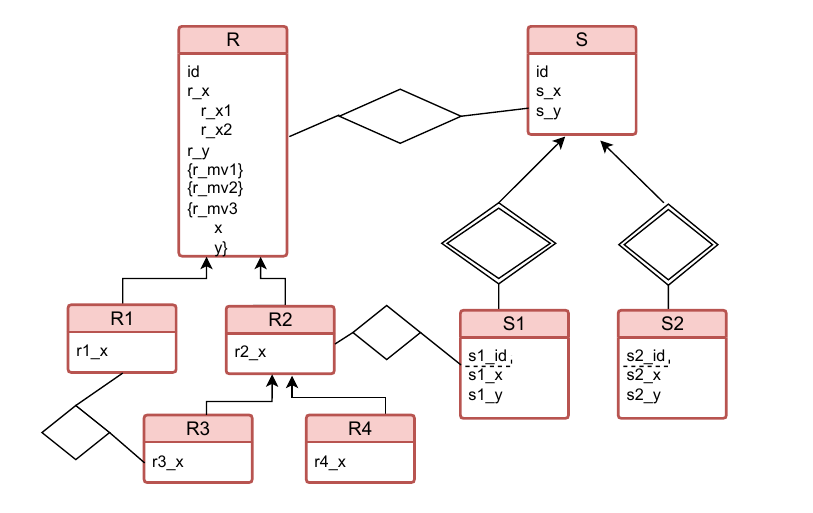}
  \caption{E/R Schema for Illustrative Experiments}
 \label{fig:er_synthetic_example}
\end{figure}

\section{Illustrative Experiments}
\label{sec:experiments}
We present a set of illustrative experiments to demonstrate the benefits of the logical independence afforded by the E/R model, 
and to discuss some of the opportunities for future research.  
We use a synthetic E/R schema as shown in Figure \ref{fig:er_synthetic_example}, consisting of 8 entity sets, including a type 
hierarchy consisting of 5 entity sets, and two weak entity sets. We consider a few different mappings of these to the underlying
relational database (PostgreSQL):
\begin{itemize}
\item (M1) Fully normalized, with separate tables for the multi-valued attributes and a separate table for each subclass consisting only of the attributes unique to that subclass;  
\item (M2) The three multi-valued attributes stored using PostgreSQL array data types;
\item (M3) The type hierarchy mapped to a single relation, with a special {\em type} attribute; 
\item (M4) The type hierarchy mapped to 5 disjoint relations; 
\item (M5) The two weak entity sets folded into $S$ using custom composite types; and
\item (M6) Multi-relational representation where R2 and S1 are joined and stored in a single table.
\end{itemize}

As we illustrate below, there are significant quantitative differences between these mappings, but in today's systems, the choice has to be made pretty early in the development 
cycle and is hard to change. 
There are also important qualitative differences
between the mappings. For instance, in M3, where $R2$ is not a separate relation, any constraints on the relationships between 
$R2$ and $S1$, or between $R1$ and $R3$ would be difficult to enforce. It's an interesting research question as to how to quantify such differences. 

We ran a series of experiments against these six mappings and a synthetically generated database containing approximate
5,000,000 entries in total. We discuss a few of the results; however, we note that many of the results are along the lines
of what one would expect given the schemas above. All queries were run 10 times, and the median time is reported.

We compared the performance of M1 and M2 on a simple query that outputs the three multi-valued attributes for all the $R$ entities. For M1,
this requires a multi-way join, resulting in a 22x performance difference between the two (M1 = 66.42s vs M2 = 2.88s), whereas a query that 
asks just for all the values of \verb|r_mv1| is about 30\% faster on M1 (M1 = 0.39s vs M2 = .5s), representing the cost of unnesting. Surprisingly a 
query that asks for the \verb|r_mv1| values given 
a \verb|r_id| showed a 145x performance difference (M1 = 40ms vs M2 = 0.3ms), likely due to it not being able to use an index on M1 (\verb|r_id| is a key
for M2, but not for M1). Especially in the context of reactive applications, the latency differences may be highly significant.
On the other hand, a query that looks for intersection of \verb|r_mv1| and \verb|r_mv2| across all tuples runs about 3.6x faster with M1 than with M2 (M1 = 0.63s vs M2 = 2.29s), 
because of the unnesting overhead. 

Next, we compare the three alternative representations for the type hierarcy (M1, M3, and M4). 
For a query that simply lists all the information for the $R3$ entities, M1 (which requires a 3-way join) was about 5x slower
than M3 (M1 = 2s vs M3 = 0.4s), and M3 in turn is about 2.7x slower than M4 (although there is no join needed for either M3 or M4, the amount of data scanned is significantly smaller for M4). 

Surprisingly, for a query that joins $R$ with $S$ with predicates on both relations, M1 and M4 
performed very similarly, which is surprising given that M4 requires a 5-relation union.
However, the performance gap between these three representations significantly increases for more complex queries. The queries for M1
and M4 also get quite verbose if any reasoning across the three relations is required.

Next, we look at the effect of folding \verb|S1| and \verb|S2| as arrays of composite types inside \verb|S| (M5 vs M1). 
A query that asks for all the information across the three entities for a given set of 10000 \verb|s_id|s ran about 2.2x 
slower on M1 due to the extra joins needed there. On the other hand, any queries that require unnesting of those 
composite arrays run much slower; for example, a query that joins \verb|S1| and \verb|R| runs about 4x slower on M5 than
M1.

Finally, comparing M1 and M6, we see that a query that can utilize the pre-computed join runs significantly faster on
M6, but queries that only involve one of those two tables get more expensive. As we noted earlier, compact
multi-relation storage formats are needed to make a representation like M6 viable.

In addition to showing the trade-offs between the different representations that can be exploited through the increased logical independence, 
the experiments also highlight some of the inefficiencies of PostgreSQL that, we believe, can be addressed relatively easily. At the same
time, they also suggest that different storage layouts and specialized operators may be needed to handle complex inheritance 
hierarchies and highly nested structures.

\section{Prior Work}
\label{sec:history}
There is a long history of work on richer data models, to provide better support for complex objects, relationships between them, and inheritance hierarchies; to display database constraints more intuitively; to provide an
easy-to-understand notation; and to make it easier to write queries against the data. We cannot do justice to that long line of work, even if we were to only focus on the work on the E/R model, but we discuss a few of those
works here. 

Several early works, including Cattell et al.~\cite{cattell1980entity, cattell1}, Elmasri and Larson~\cite{elmasri1985graphical}, Czejdo et al.~\cite{czejdo1990graphical},
        Zhang and Mendelzon~\cite{zhang1983graphical}, etc.,
developed graphical data manipulation languages for the E/R model or its variations. There were also several works that looked at non-graphical languages, generalizing relational algebra or SQL, including 
Elmasri and Wiederhold~\cite{elmasri1981gordas}, Parent and Spaccapietra~\cite{parent1984entity}, and Hohenstein and Engels~\cite{hohenstein1992sql}. There is also a long line of work on query languages for non-1NF relational databases, e.g., Roth et al.~\cite{roth1987sql,roth1988extended} and more recently, Carey et al.~\cite{carey2024sql++}. 

From implementation perspective, the early work on object-oriented databases, such as Exodus~\cite{carey1988exodus}, supported complex object types as well as object-oriented features like inheritance and aggregation, encompassing the E/R model (although much of that work does not explicitly focus on the E/R model). Microsoft's ADO.NET Entity framework, and the Entity SQL language, form perhaps the most well-known modern example of this approach~\cite{castro2007ado,melnik2008compiling,rull2013query}. That line of work has looked at a number of different implementation aspects, including bidirectional views for data transformations, query containment, and query optimization.

\section{Conclusion and Future Work}
This work is motivated by the observation that, although there has been an immense amount of work and progress on more
efficient physical implementations including different types of data storage representations and query execution algorithms, 
the user-facing aspects of databases have largely remained unchanged. The gap has been filled by the industry in coming up 
with different abstractions and layers, often on top of a relational database system. As we illustrated in this paper, this 
has significant performance implications, but more importantly, there are a number of data governance and usability implications 
because of the low level abstractions that continue to be used by the most popular database systems. We believe that E/R 
model represents a very good compromise that allows us to raise the abstraction level sufficiently without losing the benefits 
of simplicity. By introducing the higher level of logical independence inside the database itself, it opens up many more optimization 
opportunities that are worth researching further, especially in an era where machine learning techniques are poised to 
fundamentally change many aspects of programming and application development. Some of the research challenges we are
planning to investigate include automatically identifying the best physical representations for a given schema
and workload, more efficient graph-like columnar storage representations, partitioning techniques to handle
distributed and parallel environments~\cite{smith2020scalable}, optimizing API-driven workloads (that often contain
nested outputs), and schema evolution and versioning.

\section{Acknowledgements}
We would like to thank Michael Carey and Phil Bernstein for their encouragement and for providing references to closely related prior work, 
and to the anonymous reviewers for their feedback and constructive suggestions.

\bibliographystyle{ACM-Reference-Format}
\bibliography{main,new_refs}

\end{document}